\documentclass[preprint]{elsarticle}
\usepackage{amsmath,bm,amsfonts}
\newcommand{\ph}{\bm{\phi}}
\usepackage{dsfont}

\begin{document}
\begin{frontmatter}
\title{Lifshitz tricritical point and its relation to the FFLO superconducting state}
\author[snb]{Arghya Dutta \corref{cor}}
\ead{arghya@bose.res.in}
\author[snb,hri]{Jayanta K. Bhattacharjee}
\address[snb]{S. N. Bose National Centre for Basic Sciences, Block-JD, Sector-3, Salt Lake, Kolkata, Pin-700098, India.}
\address[hri]{Harish-Chandra Research Institute, Chhatnag Road, Jhusi,
Allahabad-211019, India.}
\cortext[cor]{Corresponding author. Tel: +91 (033) 2335 5706 fax:
+91 (033) 2335 3477}

\begin{abstract}
We study the phase diagram of spatially inhomogeneous Fulde-Ferrell-Larkin-Ovchinnikov(FFLO) superconducting state using the Ginzburg-Landau(GL) free energy, derived from the microscopic Hamiltonian of the system, and notice that it has a very clear Lifshitz tricritical point. We find the specific heat jumps abruptly near the first-order line in the emergent phase diagram which is very similar to the recent experimental observation in layered organic superconductor. Comparison with experimental data allows us to obtain quantitative relations between the parameters of phenomenological free energy. The region of the phase diagram where the specific heat jumps can be probed by doing a dynamical analysis of the free energy.
\end{abstract}
\end{frontmatter}

Almost 50 years ago Fulde and Ferrell\cite{ff} and Larkin and Ovchinikov\cite{lo} made a startling prediction about the possibility of a spatially inhomogeneous superconducting state at high magnetic field and low temperatures. The new-proposed FFLO state was unique as it had a spatially-modulated order parameter, in contrast to the spatially-homogeneous order parameter of a standard Bardeen-Cooper-Schrieffer(BCS) superconducting state. The FFLO state, other than being theoretically interesting, has major technological importance due to its high superconducting current densities and thus remains a very active field in both theoretical and experimental research till date.

The FFLO state, from a theoretical perspective, is  quite intriguing as it retains its superconducting property overcoming the orbital and Pauli-paramagnetic pair-breaking effects, even at very high magnetic fields. FFLO state has been vigorously studied for different physical systems\cite{rajagopal01, matsuda07,agterberg09,liao10, leo11,torma11, cai11} which vary from heavy-fermionic superconductors to dense quark matter. The experimental search for the FFLO superconductor is all the more interesting as, due to the stringent experimental conditions needed, the conclusive evidence of FFLO state is still lacking. The FFLO state will only occur if the superconductor is in the clean limit, i.e. the mean free path is much larger than the coherence length and the Maki parameter, a measure of the relative strength of orbital and Zeeman coupling to the external magnetic field, $\alpha=\sqrt{2}H_{orb}/H_P$ is greater than 1.8. The most promising candidates obeying these conditions are the heavy-fermionic superconductor $\text{CeCoIn}_{\text{5}}$ \cite{bianchi03,young07,kenzelmann08,koutroulakis10,kumagai11} and layered, or quasi two-dimensional(2D), organic superconductors.

The 2D organic superconductors have become the focus of late as they are, in most cases, clean superconductors and the orbital pair-breaking effect is mostly suppressed with a magnetic field applied parallel to the conducting layers. In an experiment done on the 2D organic superconductor $\kappa$-%
(BEDT-TTF)$_2$Cu(NCS)$_2$, in which BEDT-TTF is bisethylenedithio-%
tetrathiafulvalene, Lortz et al\cite{lortz07} first provided a true thermodynamic evidence for an increase of the upper critical field and showed the existence of a narrow intermediate superconducting region, which they claimed to be a FFLO state. Recent magnetic torque \cite{bergk11} and nuclear magnetic resonance\cite{wright11} measurements also support their view. Interestingly, one of the most prominent features marking the onset of the FFLO state is an anomalous jump in the specific heat\cite{lortz07,beyer12}. Also in the heavy-fermionic superconductor $\text{CeCoIn}_{\text{5}}$, the FFLO state appears with an anomalous jump in the specific heat\cite{bianchi03}. There are some parallel works in cuprate superconductors\cite{Kru1,krubook} and in nanoscale superconductors\cite{kru2,kru3}. Though some very recent papers by Buzdin et al\cite{croitoru12prb,croitoru12prl} report a timely study on anisotropic effects in 2D superconductors, the specific heat anomaly remains largely unexplored.      

In this Letter we present a theoretical study of the FFLO phase diagram with the help of GL free energy, derived from the microscopic Hamiltonian\cite{casal04}. We calculate the specific heat to show
that, indeed, the onset of the FFLO phase is accompanied
by a jump in the specific heat across a first-order line.  This jump is a direct consequence of the invariable presence of a Lifshitz tricritical point(LTP) in the FFLO phase diagram. The LTP occurs as a result of keeping the relevant\cite{aharony87} higher-order gradient terms in the GL free energy. The region in the phase diagram where the discontinuity in specific heat occurs can also be experimentally ascertained from a study of the temporal behaviour of dynamical structure factor. We have shown this by suddenly quenching the system from the normal to the FFLO state across the LTP and calculating the structure factor using the time-dependent Ginzburg-Landau(TDGL) formalism.
We begin by considering the action of a two-component, spin-imbalanced Fermi gas
\begin{eqnarray}
\label{H}
S[\psi^\dagger,\psi] &=&\int d\tau d^dx\psi^{\dagger}\Bigg\lbrace \Bigg( \partial_\tau -\frac{\nabla^2}{2m}+\mu\Bigg)\mathds{1}+\delta\mu\sigma_3\Bigg\rbrace \psi+ g\int d\tau d^dx(\psi^\dagger\psi)^2.\nonumber\\   
\end{eqnarray}
$\psi=(\psi_u,\psi_d)^T$ is a column matrix containing the annihilation operators of spin-up and spin-down fermions, respectively, and $\tau$ is the imaginary time. As we are looking for the finite-temperature properties of this system, we need to use finite-temperature(or imaginary-time) Matsubara formalism. $\mu+\delta\mu$ and $\mu-\delta\mu $ are the chemical potentials of the up- and down-spin fermions, respectively. Clearly, $\delta\mu$ is the measure of imbalance. In a superconducting material this spin polarization results from the Zeeman coupling of the electron spin to a magnetic field, as discussed earlier. In ultra-cold, imbalanced Fermi gases mismatch in spin polarization causes this chemical potential imbalance. $``g"$ is the attractive electron-phonon coupling constant for 2D superconductors and for an imbalanced ultra-cold atomic system, $``g"$ can be engineered to have its desired value using Feshbach resonance\cite{kohler06}.

Starting from the action of Eq.(\ref{H}), the phenomenological GL free energy had been derived by Buzdin et al\cite{buzdin97} and Combescot et al\cite{mora03}. In this Letter, we will follow the notation of Casalbuoni et al\cite{casal04}. The GL free energy functional of a FFLO superconductor in terms of the order parameter $\ph(\mathbf{k})$ , in Fourier space, is 
\begin{eqnarray}
\Omega &=& \sum_{\mathbf{k}}\Bigg(\alpha+\frac{2\beta}{3}k^2+\frac{8\gamma}{15}k^4 \Bigg)|\ph_{\mathbf{k}}|^2+ \frac{1}{2}\sum_{\mathbf{k_i}}\bigg(\beta+\frac{4\gamma}{9}\bigg(\mathbf{k_1}^2+\mathbf{k_2}^2+\mathbf{k_3}^2+\mathbf{k_4}^2\nonumber\\
&+& \mathbf{k_1}\cdot\mathbf{k_3}
+ \mathbf{k_2}\cdot\mathbf{k_4}\bigg)\bigg) \ph_{\mathbf{k_1}}\ph^{\ast}_{\mathbf{k_2}}\ph_{\mathbf{k_3}}\ph^{\ast}_{\mathbf{k_4}}
\delta _{\mathbf{k_1}-\mathbf{k_2}+\mathbf{k_3}-\mathbf{k_4}} +\frac{\gamma}{3} \sum_{\mathbf{k_i}}\ph_{\mathbf{k_1}}\ph^{\ast}_{\mathbf{k_2}}\ph_{\mathbf{k_3}}
\ph^{\ast}_{\mathbf{k_4}}\nonumber\\
&&\ph_{\mathbf{k_5}}\ph^{\ast}_{\mathbf{k_6}}
\delta_{\mathbf{k_1}-\mathbf{k_2}+\mathbf{k_3}-\mathbf{k_4}+\mathbf{k_5}-\mathbf{k_6}},
\label{free}
\end{eqnarray}
in which $\alpha=\rho[\text{ln}(4\pi T/\ph_0)+\text{Re}[\psi(1/2 +i\delta\mu/2\pi T)]]$, $
\beta = -(\rho/16\pi^2 T^2)\text{Re}[\psi^{(2)}(1/2+i\delta\mu/2\pi T)]$ and $
\gamma = \rho/(1024\pi^4 T^4)\text{Re}[\psi^{(4)}(1/2+i\delta\mu/2\pi T)]
$. $\rho  (=p_F^2/\pi^2 v_F)$ is the density of states of
the electrons at the Fermi surface, $T$ is the temperature, $\ph_0$ is the 
superconducting gap at zero temperature and zero imbalance,
$\psi(z)\; \text{and}\; \psi^{(n)}(z) $ are the digamma function 
and $n$th derivative of the digamma function, respectively. In physical terms, the coefficients $\alpha$, $\beta$ and $\gamma$ are functions of temperature and the external magnetic field which causes the chemical potential imbalance $\delta\mu$.

We note that in Eq.(\ref{free}) there is a fourth-order gradient term (quartic in momentum) which is quadratic in order parameter. This term is essential for the occurrence of a Lifshitz point. What is vital is that the coefficient of the gradient-square, quadratic $\ph$ term and quartic $\ph$ term in Eq.(\ref{free}) changes sign at the same point, namely at $\beta=0$. This is the LTP which we will take as the hallmark of the GL free energy of Eq.(\ref{free}).

In a general study of tricritical and Lifshitz kind of behaviour, Aharony et al\cite{aharony87} started with the free energy in $d$-dimensions
\begin{eqnarray}
\label{gl}
\Omega &=& \frac{1}{2}\int d^{d}x\bigg[ r\ph^2
+\mu(\nabla_{\alpha}\ph)^{2}+(\nabla_{\alpha}^{2}\ph)^2+(\nabla_{\beta}\ph)^{2} 
 + u\ph^4+ \lambda_{1}\ph^2
\ph.(\nabla_{\alpha}^2\ph)\nonumber\\
&&+\lambda_2\ph^2(\nabla_{\alpha}\ph)^2+w\ph^6\bigg],
\end{eqnarray}
where $\alpha$ denotes an m-dimensional space with $m\leq d$ and subsequently
$\nabla_{\alpha}=\sum_{i=1}^m \widehat{i}\nabla_i$, $\beta$ denotes $(d-m)$ 
dimensional space with $\nabla_{\beta}=\sum_{i=1}^{d-m} \widehat{i}\nabla_i$.
$\ph(\mathbf{x})$ is an $n$-component order parameter with $O(n)$ symmetry.

This system can have several multicritical points. If we assume, in the spirit of mean-field theories, that the homogeneous state is the relevant one, then the derivatives in Eq.(\ref{gl}) are zero. For the usual second-order transition (critical point) the parameter $r$ is temperature dependent, changing sign at the critical temperature. If there is a second physical variable (e.g. $He_3$ concentration in the $He_3 - He_4$ mixtures or the chemical potential imbalance as over here), the coefficient $u$ of the quartic term is a function of that variable and a line of critical points can be generated. This line ends when it meets a line of first-order transitions which can occur for negative values of $u$. This end point is the tricritical point where a line of first-order and second-order transitions meet (see Fig. 1a). This simple picture of a homogeneous state is changed if the sign of the derivative squared term in the free energy is allowed to change. Then the homogeneous state is no longer the lowest energy one if this term is negative. To prevent instability, one needs a second derivative squared term of positive sign and thus a spatially modulated state is formed that minimizes the relevant part of the free energy. Hence depending on whether the derivative squared term is positive or negative, the thermodynamic state is spatially uniform or modulated. We have two transition lines once again - a line across which a disordered state becomes ordered and spatially uniform and another line across which a disordered state becomes ordered but spatially modulated. These two lines meet at the Lifshitz point (Fig. 1b). If it so happens that magically the tricritical and Lifshitz point occur simultaneously (as it does in Eq.(\ref{gl}) and Eq.(\ref{glf})) where $r$ and $u$ change sign at the same point, then one has a Lifshitz tricritical point.

The GL free energy of FFLO state in Eq.(\ref{free}) can be rewritten to show its equivalence with the Lifshitz tricritical free energy as
\begin{eqnarray}
\label{glf}
\Omega &=& \int d^{d}x\bigg[ \alpha\ph^2
+\frac{2\beta}{3}(\nabla\ph)^{2}+\frac{8\gamma}{15}(\nabla^{2}\ph)^2+\frac{\beta}{2}(\ph^2)^2
- \frac{8\gamma}{9}\ph^2
\ph\cdot(\nabla^2\ph)\nonumber\\
&-&\frac{4\gamma}{9}\ph^2(\nabla\ph)^2+\frac{\gamma}{3}(\ph^2)^3\bigg].
\end{eqnarray}
The form of the GL free energy in Eq.(\ref{glf}) will be needed while doing the mean-field analysis. Interestingly, this equation also tells us that the FFLO state belong to the $(d-d)$, or isotropic, Lifshitz-tricritical class.
\begin{figure}
\centering
\includegraphics[scale=0.14]{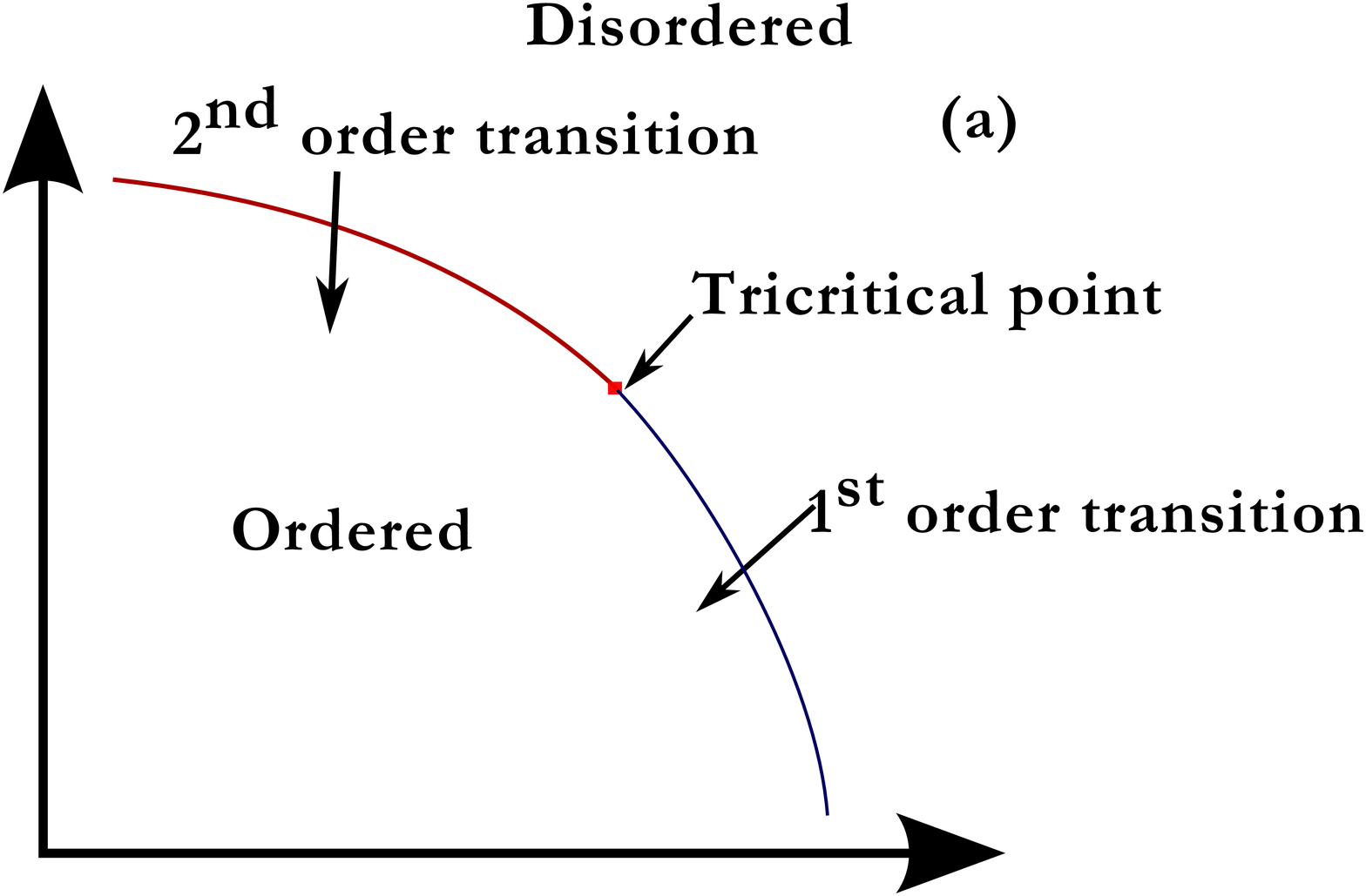}
\includegraphics[scale=0.15]{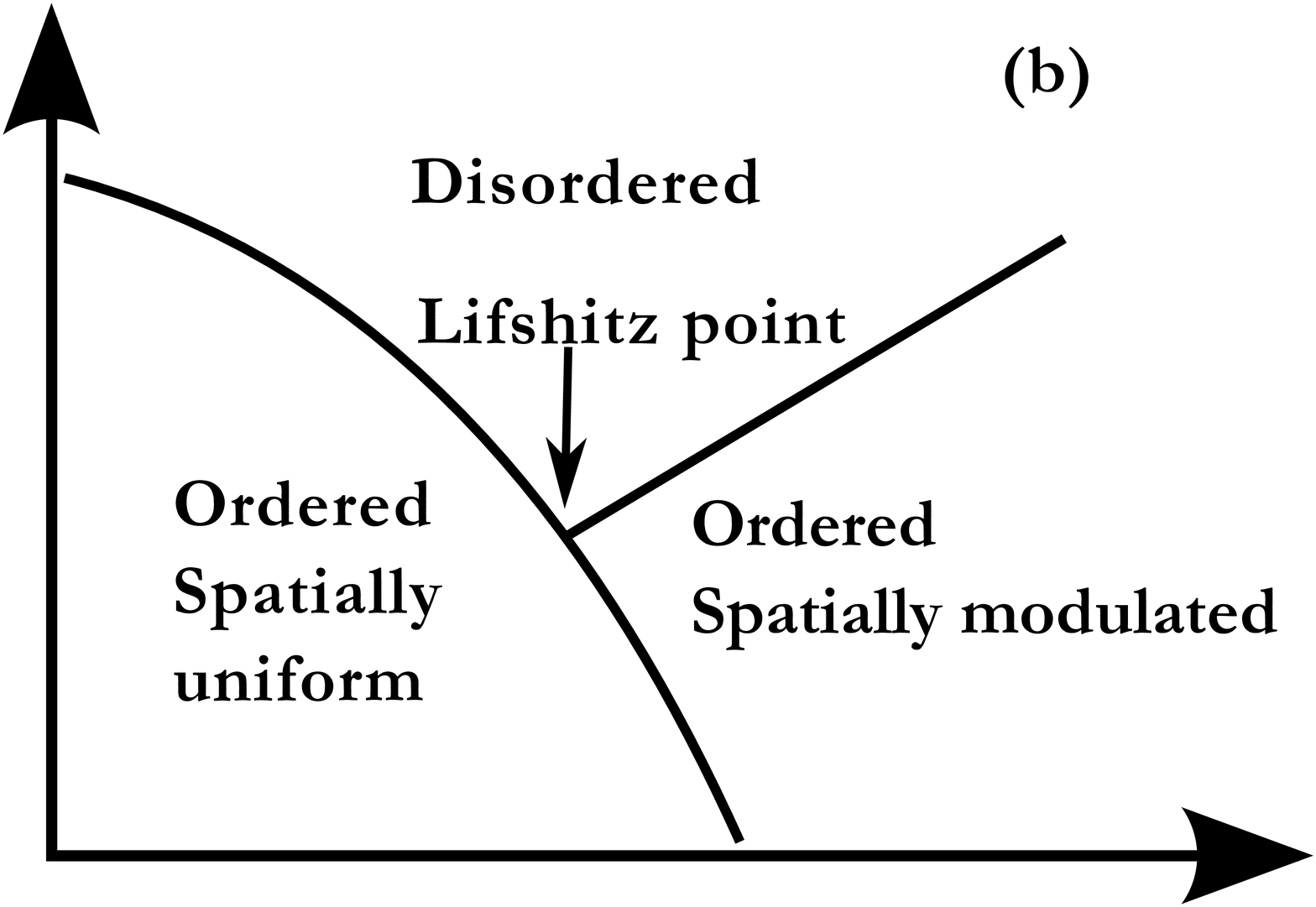}
\caption{Schematic phase diagrams showing tricritical and Lifshitz points.}
\label{multi}
\end{figure}

\begin{figure}
\centering
\includegraphics[scale=0.25]{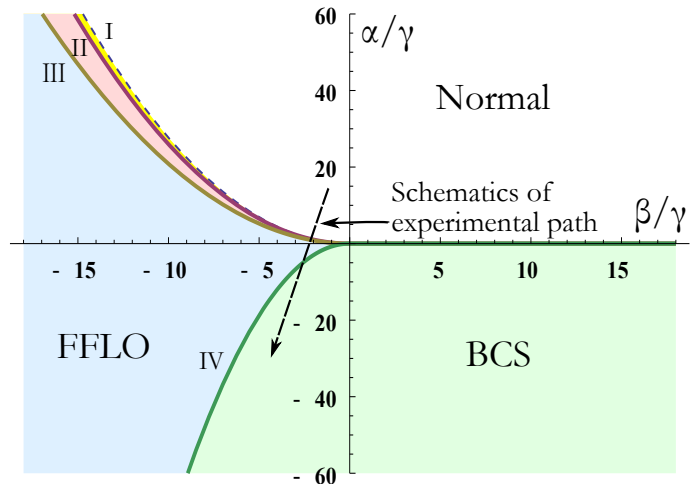}
\caption{Phase diagram of a FFLO superconductor, at mean-field level, plotted as 
a function of $\alpha/\gamma$ and $\beta/\gamma$. Graphs I and II represents the metastable and first-order line, respectively. On curve III, the coefficient of the quadratic term in order parameter vanishes. On curve IV, the FFLO state goes to the BCS state as the wave number of the order parameter goes to $0$.}
\label{pd}
\end{figure}
\begin{figure}
\centering
\includegraphics[scale=.45]{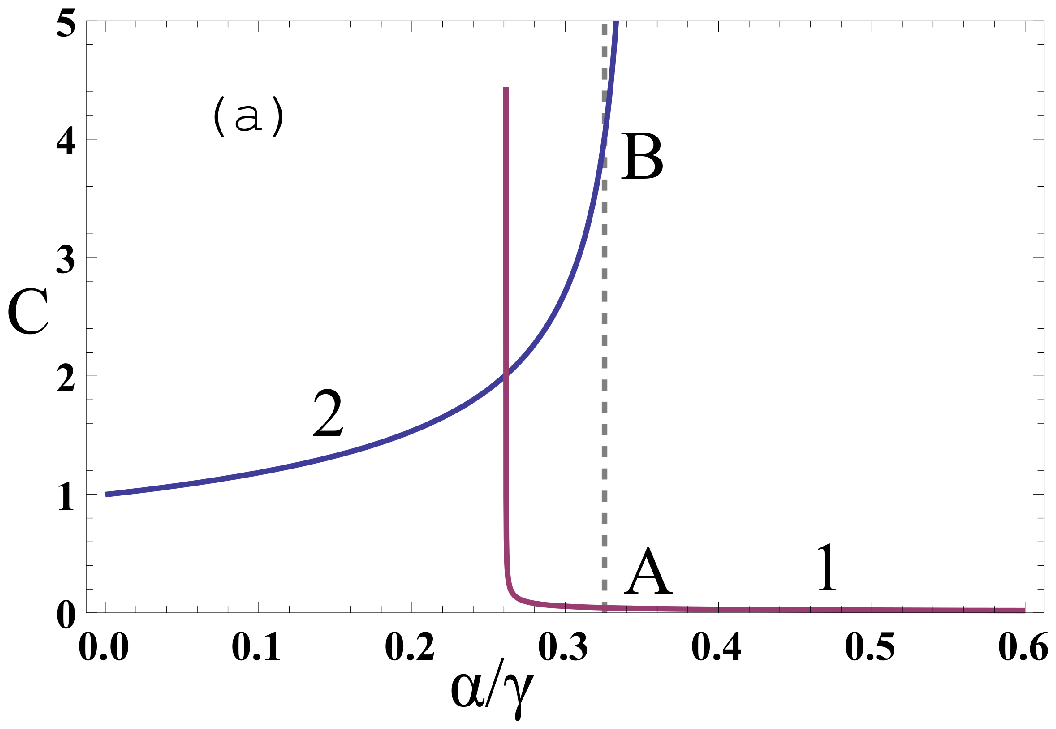}
\includegraphics[scale=.45]{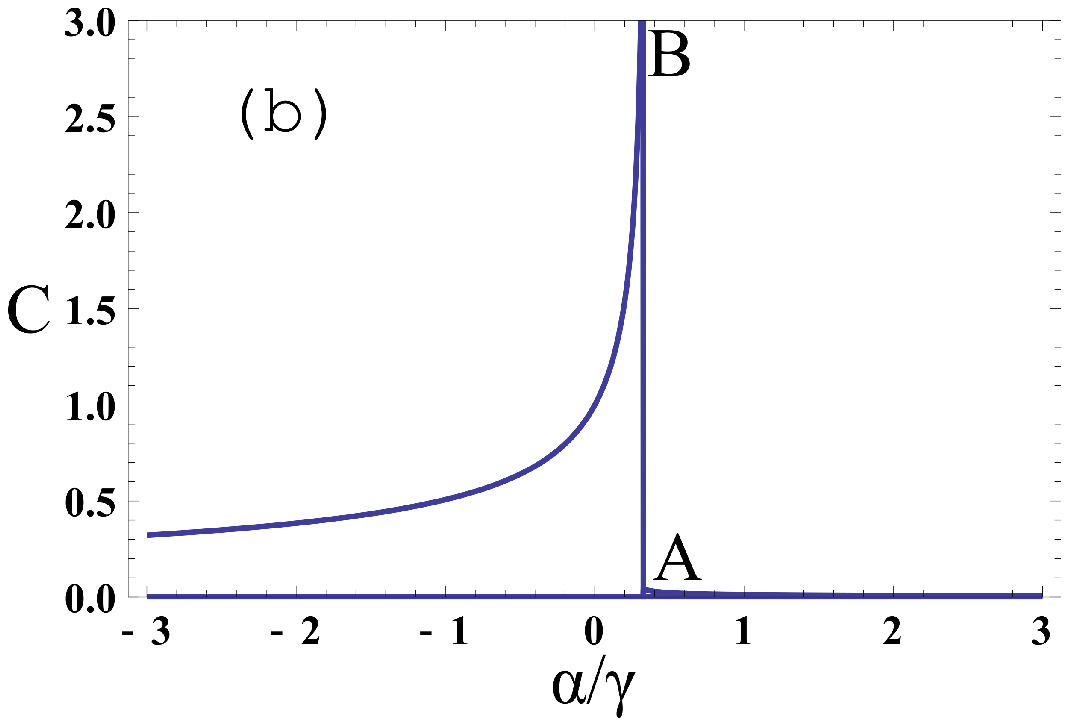}
\includegraphics[scale=.5]{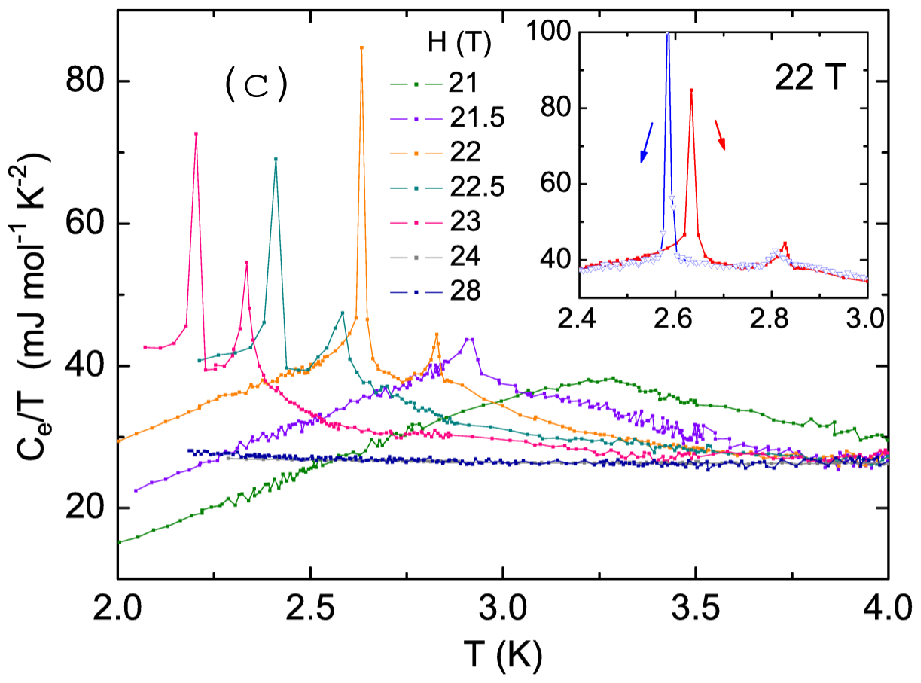}
\caption[]{Plot of the specific heat as a function of the scaled temperature. As the system is cooled from the normal state, it exhibits Gaussian specific heat, plotted as curve (1) in Fig.(a). At the first-order transition temperature, the specific heat jumps from point A to B and starts following the mean-field specific heat, plotted as curve (2). Fig.(b) focusses only on the jump in specific heat . The experimental plot of specific heat of the layered organic superconductor, as obtained by Lortz et al\cite{lortz07}, showing the anomalous jump of the specific heat is shown in Fig.(c).}
\label{spheat}
\end{figure}

We now analyse Eq.(\ref{glf}) in two-dimensions to explore the mean-field phase diagram of FFLO state. For $\beta>0$, the free energy minimum occurs at a spatially-constant $\ph$, which we call $\ph_C$. Then Eq.(\ref{glf}) reduces to $\Omega=V[\alpha\ph_C^2+\beta\ph_C^4/2+\gamma\ph_C^6/3]$. The term proportional to $\ph_C^6$ is redundant (i.e. it is now a $\ph^4$ model), and, for $\alpha<0$ the system goes from the normal state $(\ph_C=0)$ to the homogeneous BCS superconducting state with zero center-of-mass momenta Cooper pairs.

For $\beta<0$ BCS state no longer remains the minimum of GL free energy and we envisage a periodic variation of $\ph$ like $\ph(\mathbf{x})=\ph_0e^{i\mathbf{k}\cdot
\mathbf{x}}$. With this $\ph(\mathbf{x})$, the GL free energy in Eq.(\ref{glf}) reduces to
\begin{eqnarray}
\frac{\Omega}{V}=\bigg(\alpha -\frac{2|\beta| }{3}k^2+\frac{8\gamma }{15}k^4\bigg)\ph_0 ^2+\bigg(-\frac{|\beta| }{2}+\frac{4\gamma }{9}k^2\bigg)\ph_0^4+\frac{\gamma }{3}\ph_0 ^6
\end{eqnarray}
, where $V$ is system's volume. Minimising this free energy with respect to $\mathbf{k}$ and substituting $\mathbf{k}$'s minimum value in above equation we obtain an effective $\ph^6$ free energy
\begin{equation}
\label{mf}
\frac{\Omega}{V}=\Bigg(\alpha -\frac{5 \beta ^2}{24 \gamma }\Bigg) \ph_0 ^2-\frac{2 |\beta|  \ph_0 ^4}{9}+\frac{13 \gamma  \ph_0 ^6}{54}.
\end{equation}
Since $\beta < 0$ in Eq.(\ref{mf}), the minimisation of Eq.(\ref{mf}) with respect to $\ph_0$ gives a local minimum of $\Omega$ at a finite $\ph_0$ when $\alpha=\alpha_1=(259\beta^2)/(936 \gamma)$, shown as graph I in Fig.(\ref{pd}). This signals the onset of a metastable state as the normal state with $\ph_0=0$ remains the global minima. When the system cooled further keeping $\beta$ and $\gamma$ fixed, a first-order transition occurs where the free energy at $\ph_0\neq0$ becomes the global minimum. This occurs at $\alpha=\alpha_2=(27\beta^2)/(104 \gamma)$, shown as graph II in Fig.(\ref{pd}). The FFLO state thus produced has a wave number which is given by  $k_C^2=5\beta/8\gamma-5\phi ^2/12$, which clearly depends on the temperature and the magnetic field.  At an even lower temperature of $\alpha=\alpha_3=(5\beta^2)/(24 \gamma)$, the coefficient of the quadratic $\ph_0$-term in Eq.(\ref{mf}) vanishes (shown as graph III in Fig.(\ref{pd})). A further lowering of the temperature causes the wave number of the FFLO order parameter to decrease and at the boundary IV this wave number goes to $0$ signalling the end of the FFLO phase. The system thereafter makes a transition to the standard BCS phase. Thus in our scheme of things, FFLO state exists between boundaries II and IV and these are the two places where a specific heat anomaly will be seen. This is what is seen in the experiment of Lortz et al\cite{lortz07}.

With this phase diagram available, we are now ready to find the behaviour of specific heat near the first-order transition. Above the first-order transition temperature$(\alpha_2)$, the mean-field specific heat is identically zero. To calculate its value above the first-order line, we consider Gaussian fluctuations around the mean-field value and calculate the partition function $Z=\int\mathcal{D}\ph\text{exp}[-\Omega_{g}/k_BT_C]$ in which $\Omega_{g}=\sum_{\mathbf{k}}\big(\alpha-\frac{2|\beta|}{3}k^2+\frac{8\gamma}{15}k^4 \big)|\ph_{\mathbf{k}}|^2$ is the Gaussian free energy. We have considered $\beta<0$ as we are below the metastable line. We minimise this free energy with respect to $\mathbf{k}$ and write $\mathbf{k}=\mathbf{k}_C+\mathbf{l}$ where $k_C^2=5|\beta|/8\gamma$. The thermodynamic free energy per unit volume is given by $F=-k_BT_C\text{ln}Z$. The Gaussian specific heat per unit volume, the second derivative of $F$ with respect to $\alpha$, comes out to be
\begin{equation}
C_g=\frac{3\sqrt{3}k_B}{128\pi}|\beta|^{-3/2}(\alpha-5\beta^2/24\gamma)^{-1/2}.
\end{equation}
The Gaussian specific heat, evaluated at the first-order temperature, is plotted as curve $(1)$ in Fig.(3a). Here we point out that Konschelle et al\cite{konschelle07} got an anomalous exponent for $C_g$ as they considered fluctuations around $\mathbf{k}_C$ in the radial direction alone. We further observe that inclusion of the higher order $\ph$ terms will lead to a substantial increase in the rate of divergence of $C_g$ as shown by Aharony et al\cite{aharony87} who evaluated the first-order correction to the specific heat using renormalization group analysis. However, for temperatures above the first-order temperature, which is greater than the temperature at which $C_g$ diverges, the Gaussian specific heat accurately represents the fluctuation specific heat. To find the specific heat of the system below the first-order temperature, we evaluate the mean-field specific heat. This can be calculated by taking the double derivative of the  mean-field GL free energy of Eq.(\ref{mf}) with respect to $\alpha$, the temperature parameter. The mean-field specific heat per unit volume is given by
\begin{equation}
C_{mf}=18k_B(259 \beta^2 - 936 \alpha\gamma)^{-1/2}.
\end{equation}
We have plotted $C_{mf}$ at the first-order temperature as curve $(2)$ in Fig.(3a).

Let us now discuss the experimental implications of our calculation. When one cools the imbalanced fermionic system from the normal state, one measures the Gaussian specific heat. As the system cools to the first-order transition temperature, the specific heat jumps to its mean-field value and FFLO state appears. If one cools the system further, then at a sufficiently low temperature the system goes to the BCS state as the wave number of the FFLO state becomes 0 and subsequently the specific heat jumps to the BCS mean-field value signalling this transition. We have shown a schematic experimental path in Fig.(\ref{pd}), which, we argue, is the path taken in the experiment of Lortz et al showing two jumps in the specific heat. Considering the experimental data for specific heat jump at a magnetic field, applied parallel to the superconducting layers, of 22 T, we find $(\alpha\beta)^{1/2}=0.008$. The jump of the specific heat is plotted in Fig.(3b) and we have also shown the experimental specific heat diagram from Lortz et al in Fig.(3c). Also note that, in the experiment, the size of the specific heat jump increases as the magnetic field is increased. In our model, we have checked and found that the size of specific heat jump increases with increase in the chemical potential imbalance which, being directly proportional to the applied magnetic field, is equivalent to the applied magnetic field. Thus we have shown that the static analysis of the GL free energy provides the reason for the jump in specific heat.


 Now to find a way for pinpointing the region in the phase diagram where one will find the jump in specific heat in an experiment, we have studied the dynamics of this system near LTP using TDGL formalism. This study has also been encouraged by a very recent study by Sodemann et al\cite{sodemann12} who showed that finite-momentum superfluid state occupies
a large area in the phase diagram of an imbalanced fermionic system just following a quench from the normal to the superconducting state than in the equilibrium phase diagram of the same system. In our study, we look for the growth of FFLO phase following a quench. The TDGL for this system can be written as $\partial\ph_k/\partial t=-\Gamma (\delta\Omega/\delta\ph_{-k})$, in Fourier space where $\Omega$ is the GL free energy in Eq.(\ref{free}). The quantity of experimental relevance is the dynamical correlation function or the structure factor- $S(\mathbf{k},t)=\langle
\ph_{\mathbf{k}}(t)\ph_{-\mathbf{k}}(t)\rangle$. Using the TDGL we get
\begin{eqnarray}
\label{quench}
&&\frac{\partial S_{\mathbf{k}}}{\partial t} = -4\Gamma\Bigg(\alpha+\frac{2\beta}{3}k^2+\frac{8\gamma}{15}k^4 \Bigg)S_{\mathbf{k}}- 4\beta\Gamma\Bigg\langle  \sum_{\mathbf{k_1},\mathbf{k_2}} \ph_{\mathbf{k_1}}\ph_{\mathbf{k_2}}\ph_{\mathbf{k-k_1-k_2}}\ph_{\mathbf{-k}}\Bigg\rangle \nonumber\\
&&-4\gamma\Gamma\Bigg\langle\sum_{\mathbf{k_1},\mathbf{k_2},\mathbf{k_3},\mathbf{k_4}} \ph_{\mathbf{k_1}}\ph_{\mathbf{k_2}}\ph_{\mathbf{k_3}}\ph_{\mathbf{k_4}}\ph_{\mathbf{k-k_1-k_2-k_3-k_4}}\ph_{\mathbf{-k}}\Bigg\rangle\nonumber\\
&&-\frac{16\gamma\Gamma}{9}\Bigg\langle\sum_{\mathbf{k_1},\mathbf{k_2}}(2k^2+k_1^2+2k_2^2+\mathbf{k_1}\cdot\mathbf{k_2}-\mathbf{k}\cdot\mathbf{k_1}- 3\mathbf{k}\cdot\mathbf{k_2})\ph_{\mathbf{k_1}}\ph_{\mathbf{k_2}}\ph_{\mathbf{k-k_1-k_2}}\ph_{\mathbf{-k}}\Bigg\rangle.\nonumber\\
\end{eqnarray}

To make progress, we perform the Hartree approximation, in which the correlation functions factor, to the quartic and sextic correlations. Implementing this on Eq.(\ref{quench}), we arrive at
\begin{eqnarray}
\label{qm}
\frac{\partial S_{\mathbf{k}}}{\partial t} &=& -4\Gamma\bigg[\alpha+\frac{2\beta}{3}k^2+\frac{8\gamma}{15}k^4 +3\beta\sum_{\mathbf{p}}S_{\mathbf{p}}+\frac{8\gamma}{9}k^2\sum_{\mathbf{p}}S_{\mathbf{p}}+\frac{8\gamma}{9}\sum_{\mathbf{p}}p^2S_{\mathbf{p}}\nonumber\\
&+&15\gamma\Big(\sum_{\mathbf{p}}S_{\mathbf{p}}\Big)^2\bigg]S_{\mathbf{k}}
\end{eqnarray} 
From this equation it follows that at very short time, when linear approximation holds, all modes decay for $\beta<0$ if we are in the region II of the phase diagram(Fig.(\ref{pd})). To look for the long-time dynamics we work, as before, for $\beta<0$ and expand the wave number around $\mathbf{k}_C$, i.e. write $\mathbf{k}=\mathbf{k}_C+\mathbf{l}$. Restricting to terms quadratic in `$\mathbf{l}$' we can rewrite Eq.(\ref{qm}) as 
\begin{equation}
\label{qle}
\frac{\partial S_{\mathbf{k}}}{\partial t}=-\frac{16\Gamma\beta l^2}{3}S_{\mathbf{k}}-f(t)S_{\mathbf{k}}
\end{equation}
where 
\begin{eqnarray}
f(t) = -4\Gamma\bigg[\alpha-\frac{5 \beta ^2}{24 \gamma } +\bigg(3\beta+\frac{8\gamma}{9}k^2\bigg)\sum_{\mathbf{p}}S_{\mathbf{p}}+\frac{8\gamma}{9}\sum_{\mathbf{p}}p^2S_{\mathbf{p}}
+15\gamma\Big(\sum_{\mathbf{p}}S_{\mathbf{p}}\Big)^2\bigg].
\end{eqnarray} 
This allows us to write the solution of Eq.(\ref{qle}) as 
\begin{equation}
 S_{\mathbf{k}}(t)=\Delta\text{exp}\Big[-\frac{16\Gamma\beta l^2}{3}t-g(t)\Big]
\end{equation}
where $g(t)=\int_0^t dt'f(t')$. If one now demand, as one does in this kind of self-consistent calculation\cite{bray94,basu04}, that $\sum_{\mathbf{p}}S_{\mathbf{p}}$ goes to some constant value as $t\rightarrow\infty$, thus obtaining the ordered state; then for $t\rightarrow\infty$ one can neglect the sub-leading $\sum_{\mathbf{p}}p^2S_{\mathbf{p}}$ term, and, the final result for the structure factor comes out to be
\begin{equation}
\label{lqf}
S_{\mathbf{k}}(t)=\Delta t^{3/2}\text{exp}\Big[-\frac{8\gamma}{15}(k^2-k_C^2)t\Big]
\end{equation}
This result shows that at long time after the quench $S_{\mathbf{k}}(t)\rightarrow0$ for all wave numbers expect $\mathbf{k}=\mathbf{k}_C$. So determination of the parameter $k_C^2$ actually fixes the ratio $\beta/\gamma$ of the phenomenological model of Eq.(\ref{free}).

In conclusion our analysis explains the specific heat data, the first-order transition and the exact realization of the phase diagram of the FFLO state in 2D organic superconductors. In particular, we have studied the FFLO phase diagram using the phenomenological GL free energy and found that the phase diagram contains a LTP. We have calculated the specific heat and found that it jumps near the first-order line in the phase diagram as we lower the temperature, as also found in an experiment by Lortz et al\cite{lortz07}. By doing a dynamical analysis of the GL free energy, we suggest a way to explore the portion of the phase diagram where one should look for the jump. In passing, we note that for a complete explanation of the FFLO state some more material-dependent inputs like the exact shape of the Fermi surface, the mean free path and other parameters are needed, in addition to our phenomenological model. A possible extension of our work can be performing the dynamical analysis beyond the Hartree approximation and exploring the elusive FFLO state further. 

One of the authors (A.D.) thanks Council of Scientific and Industrial Research, India for financial support in the form of fellowship (File No.09/575(0062) /2009-EMR-1) and Harish-Chandra Research Institute for hospitality and support during visit. We also thank the reviewers for pertinent comments. 


\end{document}